\documentclass[12pt, prd, amsmath, amssymb, aps, natbib]{revtex4-2}

\usepackage{bm}
\usepackage{hyperref}

\newcommand{\n}{\text{n}}
\newcommand{\A}{\text{A}}
\newcommand{\B}{\text{B}}
\newcommand{\CC}{\text{C}}
\newcommand{\D}{\text{D}}
\newcommand{\M}{\text{M}}
\newcommand{\N}{\text{N}}
\newcommand{\K}{\text{K}}
\newcommand{\LL}{\text{L}}

\def\beq{\begin{equation}}
\def\eeq{\end{equation}}
\def\bear{\begin{eqnarray}}
\def\ear{\end{eqnarray}}
\def\nn{\nonumber\\ {}} 
\def\dst{\displaystyle }

\newcommand{\Pl}{\text{\tiny{Pl}}} 

\begin{document}

\title{Black holes in f(R) theory of gravity with compact extra dimensions}


\author{Arkady A. Popov}
\email{apopov@kpfu.ru}
\affiliation{N.I. Lobachevsky Institute of Mathematics and Mechanics, 	Kazan  Federal  University, \\ 
18 Kremlyovskaya street, Kazan, 420008 Russia}

\author{Sergey G. Rubin} \email{sergeirubin@list.ru}
\affiliation{%
 National Research Nuclear University MEPhI (Moscow Engineering Physics Institute),\\ 
31 Kashirskoe shosse, Moscow, 115409 Russia
}%


\begin{abstract}
We study static, spherically symmetric solutions in $f(R)$ gravity within a $D = 4+n$-dimensional spacetime, where the extra dimensions form a compact $n$-sphere of constant radius. 
We derive an exact solution in which the four-dimensional part of the metric corresponds to the Schwarzschild–de Sitter metric, while the extra dimensions are stabilized at a constant radius $L_0$. A consistency condition relates the size of the internal space to the effective four-dimensional cosmological constant $\Lambda_4 = (n-1)/L_0^2$, which is generally too large to be compatible with observations. To overcome this issue, we include the vacuum polarization effects of quantized matter fields nonminimally coupled to curvature. Using the semiclassical approach, we obtain an asymptotically flat four-dimensional Schwarzschild solution as a limiting case, where the size of the extra sphere is determined by the vacuum expectation values of the quantum fields. Finally, we discuss the dependence of the effective four-dimensional Planck mass and the extra-dimensional radius on the radial coordinate.
\end{abstract}

\maketitle


\section{Introduction}
\label{intro}

The existence of a solution in Einstein's theory of gravity that describes a Schwarzschild black hole allows us to demonstrate that the Newtonian limit of General Relativity exists \cite{Landau_Field_Theory}. A modification of Einstein's theory of gravity is one way to explain cosmological observations, in particular the so‑called ''dark energy''.   Starting from the pioneering works of Nordstr{\"o}m, Kaluza and Klein \cite{2007physics...2221N, Kaluza:1921tu,Klein:1926tv}, the idea that our observable four-dimensional universe is embedded in a larger $\D$-dimensional manifold has become a cornerstone of modern theoretical physics, particularly within the framework of string theory and braneworld  models. The most discussed class of static spherically symmetric spaces in four dimensions are those described by Gross and Perry \cite{Gross:1983hb} and Davidson and Owen \cite{Davidson:1985zf} 
\bear   
ds^2 &=& \left( \frac{ar-1}{ar+1} \right)^{2\epsilon k} dt^2 - \frac{1}{a^4 r^4} \left( \frac{(ar+1)^{{\epsilon (k-1)+1}}}{(ar-1)^{{\epsilon (k-1)-1}}} \right)^2 \Big( d r^2 + {r}^2 d \Omega_{2}^2 \Big) 
\nn &&
-\left( \frac{ar+1}{ar-1} \right)^{2 \epsilon} d y^2 \, , \qquad \epsilon^2 (k^2-k+1) = 1 \, ,
\ear
as well as Myers-Perry black holes \cite{Myers:1986un}
\begin{equation}   
ds^2 = \left( 1- \left( \frac{r_+}{r} \right)^{D-3} \right) dt^2 -\frac{dr^2}{\dst \left( 1- \left( \frac{r_+}{r} \right)^{D-3} \right)} 
- {r}^2 d \Omega_{\D-2}^2 \, .
\end{equation}
A fundamental problem in these models is the mechanism of dimensional reduction: how a complex higher-dimensional geometry can consistently yield the well-known solutions of vacuum GR at low energies. Specifically, the Schwarzschild metric — the unique spherically symmetric vacuum solution in $4D$ — serves as the primary testbed for any modified theory of gravity.

In this paper, we consider one particular modification of Einstein’s theory of gravity: $f(R)$ gravity in 
$\D$-dimensional spacetime. We assume that the extra $\D-4$ dimensions are a compact Riemannian manifold without boundary. We further require that the characteristic size of this manifold be small enough to escape current observation, at least in the region of four‑dimensional spacetime where experiments are possible.

The aim of this paper is to demonstrate that the Newtonian limit exists in $f(R)$ gravity with compact extra dimensions of small size. For this purpose, we give an explicit example of a solution $f(R)$ of the theory of gravity in a space of D dimensions, the four-dimensional part of which describes the Schwarzschild space, and the extra subspace is a D-4-dimensional sphere of constant radius $L_0$. More definitely, our system is described by metric
\begin{equation} 
\label{sch}
ds^2 = \left(1-\frac{2M}{r} \right) dt^2 - \frac{ dr^2}{\dst \left(1-\frac{2M}{r} \right)} -r^2 \left(d\theta^2 + \sin^2 \theta d\varphi^2 \right)
- {L_0}^2 d {\Omega_n}^2.
\end{equation}

This paper is structured as follows. In Section II we construct a static spherically symmetric solution of pure f(R) gravity in the four-dimensional part of space, assuming that the extra dimensions are a $\n$-dimensional sphere of constant radius. In Section III, we solve the problem by taking into account the vacuum polarization of quantized fields. In Section IV we rewrite the results obtained in the previous section in four-dimensional units. In Section V we show that the four-dimensional Planck mass can be a function of the radial coordinate.

We use the conventions for the curvature tensor $R_{\ \M\N\K}^\LL =\partial_\K\Gamma_{\M\N}^\LL-\partial_\N \Gamma_{\M\K}^\LL +\Gamma_{\CC\K}^\LL\Gamma_{\N\M}^\CC-\Gamma_{\CC\N}^\LL \Gamma_{\M\K}^\CC$ and the Ricci tensor $R_{\M\N}=R^\K_{\ \M\K\N}$. The units $\hbar=c=1$ are used throughout the paper.

\section{Schwarzschild-de Sitter black holes with an additional n-dimensional sphere of constant radius}
 \label{f(R)}

Consider $f(R)$ gravity in a $\D = 4 + \n$-dimensional manifold $M_\D$:
\begin{equation} \label{SfR}
S =  \frac{{m_\D}^{\D-2}}{2} \int_{M_\D}  d^{\D} X \sqrt{|g_{\D}|} \, f(R) \, . 
\end{equation}
where $f(R)$ is a function of the $\D$-dimensional Ricci scalar $R$,  $m_\D$ is the $\D$-dimensional Planck mass,  $\A, \B, \CC, ... = 1, 2, 3,   \dots, \n+4 $, $g_{\D} \equiv \det g_{\A\B}$. 

Variation of the action \eqref{SfR} with respect to the metric $g^{\A\B}_\D$ leads to the known equations
\bear             \label{EE}
&&- \frac{1}{2} f(R) \delta^\B_\A + \left(R^\B_\A + \nabla_\A \nabla^\B - \delta^\B_\A \square \right)f_R= 0,
\ear
with $f_R = {df(R)}/{dR}$, $\Box= \nabla^{\A} \nabla_{\B}$\,.

The subject of our study is 4-dimensional static spherically symmetric metric metric with a compact extra subspace. The simplest example of such a subspace is the $\n$-dimensional sphere. More definitely we can consider the metric
\begin{equation}   \label{metric1}
ds^2 = h(\rho) dt^2 - a(\rho) d \rho^2 -r(\rho)^2 \left(d\theta^2 + \sin^2 \theta d\varphi^2 \right)
- {L(\rho)}^2 d \Omega_n^2.
\end{equation}
As is well known, in this case, equations \eqref{EE} reduce to four independent equations for the unknown functions $h(\rho), a(\rho), r(\rho)$, and $L(\rho)$. Replacing the radial variable $\rho \to r$ 
\begin{equation}   \label{metric2}
ds^2 = h(r) dt^2 - b(r) d r^2 -r^2 \left(d\theta^2 + \sin^2 \theta d\varphi^2 \right)
- {L(r)}^2 d \Omega_n^2.
\end{equation}
results in only three of these four equations being independent. We will show that all these equations are solvable under the assumption $L(r)= L_0 =\mbox{const}$
\begin{equation}   \label{plus33}
ds^2 = h(r) dt^2 - b(r) dr^2 -r^2 \left(d\theta^2 + \sin^2 \theta d\varphi^2 \right)
- {L_0}^2 d {\Omega_n}^2.
\end{equation}
More precisely, we will analytically show that for a sufficiently arbitrary $f(R)$ gravity without matter, there exists a solution with a constant size of extra dimensions. For a 4D observer, this solution represents the Schwarzschild-de Sitter metric.

Nontrivial equations (\ref{EE}) in metric (\ref{plus33}) are
\bear   \label{tt2_+}
&&
\frac{{R'}^2}{b} f_{RRR} + \frac{1}{b}\left[ R'' +\left( -\frac{b'}{2 b} +\frac{2}{r} \right) R' \right] f_{RR}
+\frac{1}{b}\left( -\frac{h''}{2 h} +\frac{{h'}^2}{4 h^2} +\frac{h' b'}{4 h b} 
\right. \nonumber \\ && \left.
- \frac{h'}{h r} \right) f_R  -\frac{f}{2} = 0,
\ear
\bear  \label{rhrh2_+}
&&
\left( \frac{h'}{2 h } +  \frac{2}{r} \right) \frac{R'}{b} f_{RR} + \frac{1}{b}\left( -\frac{h''}{2 h} +\frac{{h'}^2}{4 h^2} + \frac{h' b'}{4 h b} +\frac{b'}{b r} \right)f_{R} -\frac{f}{2}= 0,
\ear
\bear  \label{thth2_+}
&& \frac{{R'}^2}{b} f_{RRR} +\frac{1}{b}\left[ R'' +\left( \frac{h'}{2 h} -\frac{b'}{2 b} + \frac{1}{r} \right) R' \right]f_{RR} 
+\left[ \frac{1}{b r} \left( - \frac{h'}{2 h } +\frac{b'}{2 b }  - \frac{1}{r}  \right) 
\right. \nonumber \\ && \left.
+\frac{1}{r^2}\right] f_{R} - \frac{f}{2} =  0,
\ear
\bear   \label{psps2_+}
 && \frac{{R'}^2}{b} f_{RRR} +\frac{1}{b}
\left[ R'' +\left(\frac{h'}{2 h} -\frac{b'}{2 b} +\frac{2}{r}\right)R' \right] f_{RR} + \frac{(n-1) f_{R}}{{L_0}^2}
 -\frac{f}{2} 
 = 0,
\ear
and
\bear \label{eqR2_+}
&  &R(r) =\frac{1}{b}\left( -\frac{h''}{h} +\frac{{h'}^2}{2 h^2} +\frac{ h'b'}{2 h b} -\frac{2 h'}{h r} +\frac{2b'}{b r} -\frac{2}{r^2}\right)
 +\frac{2}{r^2} +\frac{\n(\n-1)}{{L_0}^2},
\ear
where the prime denotes the derivative with respect to $r$.

Let us calculate the following combination of equations (\ref{tt2_+}-\ref{eqR2_+}):
$\eqref{tt2_+}+\eqref{rhrh2_+} +2\cdot \eqref{thth2_+} - 3 \cdot \eqref{psps2_+} - f_R \cdot \eqref{eqR2_+}$
\begin{equation}   \label{comb5+}
\left( R -\frac{(\n-1)(\n+3)}{{L_0}^2}  \right) f_R -\frac{f}{2} = 0,
\end{equation}
From this equation it follows that
\begin{equation}
R=R_0=\text{const}.
\end{equation}
Then the equations (\ref{tt2_+}-\ref{eqR2_+}) will take the form
\bear   \label{tt2_2+}
&&
\frac{1}{b}\left( -\frac{h''}{2 h} +\frac{{h'}^2}{4 h^2} +\frac{h' b'}{4 h b} - \frac{h'}{h r} \right) f_R(R_0) 
-\frac{f(R_0)}{2}=0,
\ear
\bear  \label{rhrh2_2+}
&&
\frac{1}{b}\left( -\frac{h''}{2 h} +\frac{{h'}^2}{4 h^2} + \frac{h' b'}{4 h b} +\frac{b'}{b r} \right)f_{R}(R_0) -\frac{f(R_0)}{2}=0,
\ear
\bear  \label{thth2_2+}
&& \left[ \frac{1}{b r} \left( - \frac{h'}{2 h } +\frac{b'}{2 b }  - \frac{1}{r}  \right)  +\frac{1}{r^2}\right] f_{R}(R_0) - \frac{f(R_0)}{2}=0,
\ear
\bear   \label{psps2_2+}
 && \frac{(n-1)}{{L_0}^2} f_{R}(R_0) -\frac{f(R_0)}{2}=0,
\ear
\beq   
R_0 =  \frac{1}{b}\left( -\frac{h''}{h} +\frac{ {h'}^2 }{2 h^2} +\frac{h' b'}{2 h b} -\frac{2 h'}{h r} +\frac{2 b'}{b r} -\frac{2}{r^2} \right) +\frac{2}{r^2}  +\frac{\n(\n-1)}{{L_0}^2} 
\eeq
A consequence of the equations (\ref{tt2_2+}, \ref{rhrh2_2+}) is
\beq
\left( \frac{h'}{h} +\frac{b'}{b}\right) \frac{f_{R}(R_0)}{b \, r} =0.
\eeq
This equation has the following solution
\beq \label{bb}
b(r) = \frac{C}{h(r)},
\eeq
where $C$ is the integration constant.
This standard relationship of metric functions under the assumption of spherical symmetry reduces the system of equations under study to the following system 
\beq \label{tt22}
\left( \frac{h''}{2} +\frac{h'}{r} \right) \frac{f_{R}(R_0)}{C} +\frac{f(R_0)}{2} =0,
\eeq
\beq \label{thth22}
\left( \frac{h'}{r} +\frac{h}{r^2} -\frac{C}{r^2} \right) \frac{f_{R}(R_0)}{C } +\frac{f(R_0)}{2}  =0,
\eeq
\beq   \label{fR22}
\frac{(n-1)}{{L_0}^2} f_{R}(R_0) - \frac{f(R_0)}{2} =0,
\eeq
\beq   \label{R22}
R_0 = - \frac{1}{C}\left( h'' +\frac{4 h'}{r} + \frac{2 h}{r^2}\right) +\frac{\n(\n-1)}{{L_0}^2} +\frac{2}{r^2}.
\eeq
The solution to the equation \eqref{thth22} is
\bear \label{hh}
h(r) =  {C}\left(1 -\frac{2 M}{r} -\frac{\Lambda_4}{3} r^2\right) =\frac{C}{b(r)}, 
\ear
where
\begin{equation} \label{L4}
\Lambda_4\equiv \frac{\dst  f(R_0) }{2f_{R}(R_0)},
\end{equation}
and $M$ is the constant of integration. The expression \eqref{hh} turns the equation \eqref{tt22} into an identity, and the expressions \eqref{R22} and \eqref{fR22} are transformed to the form
\beq \label{R33}
R_0 = \frac{\n(\n-1)}{{L_0}^2} +\frac{\dst 2 f(R_0)}{f_R(R_0)}, \quad 
\frac{2 (\n-1)}{{L_0}^2} f_{R}(R_0) = f(R_0). 
\eeq
Expressions (\ref{L4}, \ref{R33}) can be rewritten as follows
\beq \label{La4}
\Lambda_4 = \frac{(\n-1)}{{L_0}^2}, \quad
L_0^2=  \frac{(\n+4)(\n-1)}{{R_0}},  
\end{equation}
where $R_0$ is determined from the equation
\begin{equation} \label{algeq}
    R_0=\frac{(\n+4)}{2} \frac{f(R_0)}{f_R(R_0)} .
\end{equation}
Thus, the expressions (\ref{hh}, \ref{La4}, \ref{algeq}) define the general static, spherically symmetric solution of equations (\ref{EE}) in the four-dimensional part of spacetime, under the assumption that the extra space is an $n$-dimensional sphere of radius $L_0$. 

Relation (\ref{La4}) shows that the experimental constraints on the size $L_0$ correspond to large values of $\Lambda_4$, which contradicts observations. Below, we will show that taking into account the vacuum fluctuations of quantized fields nonminimally coupled to curvature makes it possible to find solutions with arbitrarily small values of $\Lambda_4$.

\section{Semiclassical f(R) theory } \label{TMN}


It is well known that the vacuum fluctuations of the quantized fields contribute to the cosmological constant making it extremely high. The fine tuning is necessary to reduce it to the observable value. In this section, we take into account these fluctuations
to show that the reduction to the observable scale is possible.

In the semiclassical theory of gravity, the spacetime geometry is determined by the expectation value of the stress–energy tensor operator of the quantized matter fields
\bear             \label{EE2}
&&- \frac{1}{2} f(R) \delta^B_A + \left(R^B_A + \nabla_A \nabla^B - \delta^B_A \square \right)f_R= -\frac{1}{m_{\D}^{\D-2}} \left< T^{B}_{A} \right> \, .
\ear
A fundamental limitation of semiclassical gravity is that the effects of the quantized gravitational field are ignored. The standard approach to mitigate this issue is to operate in the limit of a large number of fields, where the gravitational contribution becomes negligible. An additional complication is that the vacuum polarization effects are determined by the topological and geometrical properties of spacetime as a whole or by the choice of quantum state in which the expectation values are taken. Consequently, determining the functional dependence of $\langle T^{\mu}_{\nu}
\rangle$ on the metric tensor in a generic spacetime poses a formidable challenge. Herein, we shall consider quantum fields residing in the zero-temperature vacuum state defined with respect to the timelike Killing vector, which is guaranteed to exist in a static spacetime \eqref{plus33}.

In certain cases, $\langle T_{\mu \nu} \rangle$  is governed by the local properties of spacetime, making it possible to determine approximately the functional dependence of the renormalized vacuum expectation value of the stress–energy tensor operator on the metric tensor. A well-known example of such a scenario is the massive field case, where $\langle T^{\mu}_{\nu}
\rangle$ can be expanded  in powers of the small parameter
      \beq \label{ml}
      \frac1{m \, l} \ll 1,
      \eeq
where $m$ denotes the mass of the quantized field and $l$ the characteristic curvature scale of the spacetime  \cite{Schwinger:1951nm, DeWitt:1975ys, 1982PhLB..115..372F, Frolov:1983ig, Frolov:1984ra, Anderson:1994hg,  Herman:1998dz,Matyjasek:1998mq, Matyjasek:1999an, Koyama:2000zu, Matyjasek:2000iy, Popov:2001kk,  Popov:2003ne}.
For fields nonminimally coupled to the spacetime curvature in the spacetime \eqref{plus33}, an effective mass arises, determined by the constants of nonminimal coupling of the fields to the spacetime curvature. This effective mass can be nonzero even for a massless field. We illustrate this by the example of a scalar field $\Phi$ whose action takes the form
\begin{equation} \label{Sf}
S =  \frac{1}{2} \int_{M_\D}  d^{\D} X \sqrt{|g_{\D}|} \,   \Big(g^{A B} (\nabla_A \Phi) (\nabla_B \Phi) -(m^2 +\xi R) \Phi^2 \Big) \, ,
\end{equation}
where $\xi$ is the coupling constant of the scalar field to the spacetime curvature. 
In the spacetime \eqref{plus33}, under the condition
\beq \label{cond33}
l(r) \gg L_0\, \gg l_{\Pl}\, ,
\eeq  
where $l(r)$ is a characteristic scale of variation of the metric
functions $h(r)$ and $b(r)$ 
\beq
R(r) = \frac{\n (\n-1)}{{L_0}^2}\Big( 1+ O\left({{L_0}^2}/{l(r)^2} \right) \Big)\, .
\eeq
This implies that, in the action \eqref{Sf}, the constant $\xi {\n (\n-1)}/{{L_0}^2}$ is added to the squared mass, and for certain values of this parameter one-loop local expressions can be computed for the components of the vacuum expectation value of the stress-energy tensor operator, analogous to the corresponding expressions for a massive field. There exists an explicit example of such calculations in the case of a spacetime with the metric 
\begin{equation} 
ds^2 = h(r) dt^2 - b(r) dr^2 
- {L_0}^2  \left(d\theta^2 + \sin^2 \theta d\varphi^2 \right) \, ,
\end{equation}
for a massless scalar field under the condition  $l(r) \gg L_0\, \gg l_{\Pl}$ \cite{Popov:2001kk}
\bear   
\left< T^{t}_{t} \right>&=& \left< T^{r}_{r} \right> = 
\frac{1}{4 \pi^2 {L_0}^4}\left\{
{\frac {3{\xi}^{2}}{8}}
-{\frac {11\xi}{96}}
+{\frac {79}{7680}}
+ \left(-{\frac {{\xi}^{2}}{2}} +{\frac {\xi}{6}}-{\frac {1}{60}} \right)
 \ln \sqrt{ \frac{8\xi -1}{4 m_{\rm DS}^2 {L_0}^2}}
\right. \nn && \left. 
+\left(
2{\xi}^{2}
-{\frac {\xi}{2}}
+\frac{1}{32}\right) \left[
I_1\left( 2 \xi-\frac14 \right)-I_2\left( 2 \xi-\frac14 \right) \right] \ + O\Big({L_0}^2/l(r)^2 \Big)
 \right\},
\ear
\bear   
\left< T^{\theta}_{\theta} \right>&=& \left< T^{\varphi}_{\varphi} \right> =  
\frac{1}{4 \pi^2 {L_0}^4}\left\{
\left( -\frac{\xi^2}{8} +\frac{\xi}{32}-\frac{1}{512} \right)
+\left(\frac{\xi^2}{2} -\frac{\xi}{6} +\frac{1}{60} \right) \ln \sqrt{ \frac{8\xi -1}{4 m_{\rm DS}^2 {L_0}^2}}
\right.\nn &&\left.
+\left(-2 \xi^2+\frac{\xi}{2}-\frac{1}{32} \right)
\left[
I_1\left( 2 \xi-\frac14 \right)-I_2\left( 2 \xi-\frac14 \right) \right] \ + O\Big({L_0}^2/l(r)^2 \Big)
\right\}\, ,
\ear 
where
\beq
I_n(\mu)=\int \nolimits_{0}^{\infty}\frac{x^{2n-1}\ln|1-x^2|}
{1+e^{2\pi \mu x}}dx\, ,
\eeq
and $m_{\rm DS}$ is an arbitrary parameter due to the infrared cutoff. We note that, up to small terms of order  $O\Big( {L_0}^2/l(r)^2 \Big)$ these expressions define $\left< T^{B}_{A} \right>$ in the spacetime
\begin{equation} 
ds^2 =  dt^2 - dr^2 
- {L_0}^2  \left(d\theta^2 + \sin^2 \theta d\varphi^2 \right) \, .
\end{equation}
It is easy to see that the algebraic structure of the vacuum expectation value of the stress-energy tensor operator for a scalar field nonminimally coupled to curvature in the spacetime
\begin{equation} 
ds^2 =  dt^2 - dx^2 - dy^2 - dz^2
- {L_0}^2  d {\Omega_n}^2 \, 
\end{equation}
takes the form
\beq
\left< T^{B}_{A} \right> = \frac{1}{{L_0}^{4+n}} \mbox{diag} \Big(K^{t}_{t}, K^{t}_{t}, K^{t}_{t}, K^{t}_{t}, K^{5}_{5}, \dots, K^{5}_{5} \Big),
\eeq
and remains unchanged upon passing to spherical coordinates of the three-dimensional space
\begin{equation} \label{spher}
ds^2 =  dt^2 - dr^2 - r^2\Big(d\theta^2 +\sin^2(\theta) d\varphi^2 \Big)
- {L_0}^2  d {\Omega_n}^2 \, .
\end{equation}
For other fields nonminimally coupled to the spacetime curvature, the algebraic structure of the energy-momentum tensor in the spacetime \eqref{spher} is analogous. Thus, in the spacetime \eqref{plus33}, the vacuum expectation value of the energy-momentum tensor for quantized fields takes the form
\beq \label{TKr}
\left< T^{B}_{A} \right>=\sum_{k=1}^N \Big< \mathrel{\mathop{T^{B}_{A}}\limits^{\hskip-2mm (k)}} \Big> = \frac{K^{B}_{A}}{{L_0}^{4+n}}  \left( 1+ O\Big({L_0}^2/l(r)^2 \Big) \right),
 \eeq
where 
 \beq
K^{B}_{A} =\mbox{diag}\Big(K^{t}_{t}, K^{t}_{t}, K^{t}_{t}, K^{t}_{t}, K^{5}_{5}, \dots, K^{5}_{5} \Big)
 \eeq
$K^{B}_{A}$ are dimensionless constants, and $N$ is the number of fields excluding the gravitational one. In what follows, we will use these expressions to search for static, spherically symmetric solutions of equations \eqref{EE2} that are asymptotically flat in the four-dimensional part of the spacetime. 

Solutions of equations \eqref{EE2} under assumptions \eqref{plus33} and \eqref{cond33}, analogous to those presented in  Section \ref{f(R)}, take the form
\bear \label{hh2}
h(r) = \frac{C}{b(r)} = C\left(1 -\frac{2 M}{r} -\frac{\Lambda_4}{3} r^2\right), 
\ear
\begin{equation} \label{L42}
\Lambda_4\equiv \frac{\dst \Big( f(R_0) - \frac{2}{m_{\D}^{2+\n}} \frac{K^t_t}{{L_0}^{4+\n}} \Big)}{2f_{R}(R_0)}, \quad
R_0 = \frac{\n(\n-1)}{{L_0}^2} +\frac{\dst 2 \Big(f(R_0) - \frac{2}{m_{\D}^{2+\n}} \frac{K^t_t}{{L_0}^{4+\n}} \Big)}{f_R(R_0)}.
\end{equation}
Here, the quantities $f_R(R_0), f(R_0),$ and  $L_0$ are related by
\beq   \label{fR22_2}
\frac{2 (\n-1)}{{L_0}^2} f_{R}(R_0) = f(R_0)    - \frac{2}{m_{\D}^{2+\n}} \frac{K^5_5}{{L_0}^{4+\n} },
\eeq
As in the case $\left< T^{B}_{A} \right> =0$, this solution is the Schwarzschild–de Sitter solution in the four-dimensional part of the spacetime. 

We note that in Einstein's theory of gravity in four-dimensional spacetime ($f(R)=R-2 \Lambda, \ \n=0$), an asymptotically flat Schwarzschild solution is possible only when $\Lambda =0$. In the semiclassical $f(R)$
theory of gravity under consideration, the following expressions correspond to an asymptotically flat four-dimensional part of the spacetime
\beq \label{hh3}
h(r) = \frac{1}{b(r)} = \left(1 -\frac{2 M}{r}\right),
\eeq
\beq  \label{con12}
\Lambda_4=0 \  \to \  f(R_0) = \frac{2 K^t_t }{m_{\D}^{2+\n}} \left( \frac{ {R_0} }{ \n (\n-1)}\right)^{(4+\n)/2},
\eeq
\beq \label{con22}
{L_0}^2 = \frac{\n(\n-1)}{R_0}, \quad (\n-1)^{2+\n/2} f_{R}(R_0) = \frac{ \Big( K^t_t -K^5_5 \Big)}{m_{\D}^{2+\n}}  \frac{ R_0^{1+\n/2} }{ \n^{1+\n/2} }.
\eeq

As an example, we write out these expressions explicitly for the case $f(R) =aR^2 +R +c$ and $\n=2$
\beq
{R_0}=\frac{2}{{L_0}^2} , \quad 
L_0^2 \ {m_6}^2 = \frac{ K^5_5 -K^t_t }{ 2 a {m_6}^2 \pm \sqrt{4 a^2 {m_6}^4 +K^t_t -K^5_5 } },
\end{equation}
\bear \label{c1}
\frac{c}{{m_6}^2} &=& \frac{32(K^t_t +K^5_5)} {(K^t_t -K^5_5)^3} a^3 {m_6}^6 \mp \frac{16 (K^t_t +K^5_5) \sqrt{4 a^2 {m_6}^4 +K^t_t -K^5_5}}{(K^t_t -K^5_5)^3} a^2 {m_6}^4 \nonumber \\
&& +\frac{4({K^t_t}^2 +K^t_t K^5_5 -2{K^5_5}^2)}{(K^t_t -K^5_5)^3} a {m_6}^2 \mp \frac{ 2 K^5_5 \sqrt{4 a^2 {m_6}^4 +K^t_t -K^5_5} }{(K^t_t -K^5_5)^2} \,.
\ear
We note that for $a=0$ (and hence $f(R)=R+c, \ \n=2$), the curvature radius $L_0$ of the extra dimensions and the 6-dimensional cosmological constant $c$ are related solely by the vacuum polarization effect of the quantized fields.
\begin{equation} \label{L0n2}
h(r) = \frac{1}{b(r)} = 1 -\frac{2 M}{r}\, ,  \quad L_0^2 = \frac{ \sqrt{K^t_t -K^5_5 }  }{ {m_6}^2 }\, ,
 \quad
c = {m_6}^2\frac{ 2 K^5_5  }{(K^t_t -K^5_5)^{3/2}} 
\end{equation}
The second expression bears a formal resemblance to the cosmological constant term, which in standard quantum field theory results from a strong fine‑tuning of the vacuum energy contributions.

\section{Reduction to four-dimensional units}

To estimate the physical effects of our model, we must convert the results into standard four-dimensional units. Up to now, we have used the $D$-dimensional Planck mass as the unit. Let us relate it to the observable 4D Planck mass.

We begin with the general metric ansatz
\begin{equation}\label{interval}
ds^2 = g^{(4)}_{\mu\nu}(x)dx^{\mu}dx^{\nu} + g^{(n)}_{ab}(y)dy^a dy^b \, .
\end{equation}
We assume that the compact extra-dimensional space is small and highly curved compared to the large four-dimensional dimensions, such that the following inequality for the Ricci scalars holds:
\begin{equation}\label{R4n}
R_{4+n} = R_4(x) + R_n(y), \quad R_4(x) \ll R_n(y) \, .
\end{equation}

Following \cite{Bronnikov:2005iz}, we substitute \eqref{R4n} into the initial action \eqref{SfR} and perform a Taylor expansion of the function $f(R)$ around the background curvature $R_n(y)$:
\begin{eqnarray}\label{SfR2}
S &=&\int_{M_D} d^{D} X \sqrt{|g_{D}|} \,  \frac{m_{D}^{D-2}}{2} f\Big(R_4(x)+R_n(y)\Big) \nonumber \\
& =& \int d^{4} x d^n y \sqrt{|g^{(4)}g^{(n)}|} \, \frac{m_{D}^{D-2}}{2} \Big( f(R_n) + f_R(R_n)R_4(x) + O(R_4^2) \Big) \nonumber \\
& \simeq& \frac{m_4^2}{2} \int_{M_4} d^4x \sqrt{|g^{(4)}|} \Big(  R_4(x)  + O(R_4^2) \Big)
\end{eqnarray}
where $f_R(R) = \partial f / \partial R$. The third line is the standard Einstein-Hilbert action.
The term $f(R_n)$ is accommodated into the $\Lambda_4$ term 
$$\Lambda_4={\dst \frac12 \Big( f(R_0)}{\dst - \frac{2}{m_{D}^{2+n}} \frac{K^t_t}{{L_0}^{4+n}} \Big)}/{\dst f_R(R_0)}$$ 
as shown above, see \eqref{L42}.
In the solution under consideration (\ref{hh3}-\ref{con22}), 
$\Lambda_4=0$
accounts for the vacuum fluctuations of quantized fields.

By comparing the second and third lines in \eqref{SfR2}, we identify the effective 4D Planck mass squared $m_4^2$ as:
\begin{equation}
m_4^2 = m_{D}^{D-2} \int d^n y \sqrt{|g^{(n)}|} f_R(R_n) = m_{D}^{D-2} f_R(R_0) \frac{2 \pi^{(n+1)/2}}{\Gamma\Big((n+1)/2\Big)} \, .
\end{equation}
This relation allows one to calibrate the fundamental $D$-dimensional mass scale $m_D$ in terms of the observed gravitational constant $G_N = 1/m_4^2$.
Applied to the case of linear gravity considered above, $ f(R)=R+c$ and $n=2$, this yields the Planck mass in the form
\beq \label{m4}
{m_4}^2={m_{6}}^{4}\int d^2 y \sqrt{|g^{(2)}|} =4\pi {L_0}^2 {m_{6}}^{4} \, .
\eeq
To obtain a numerical estimate, let us assume that $L_0=100/m_6$. Then $m_6\sim 10^{-2}m_4\simeq 10^{-17}$~GeV.

When expressed in terms of four-dimensional scales, the solution (\ref{L0n2}) reduces to the standard four-dimensional Schwarzschild metric
\begin{equation}
h(r) = \frac{1}{b(r)} = 1 - \frac{2M}{r},
\end{equation}
in spite of the presence of extra spatial dimensions. The observed smallness of the cosmological constant $\Lambda_4$ is achieved by a fine-tuning — specifically, through the compensation of the bare parameter $c$ by the quantum corrections discussed above:
\begin{equation}
c = m_4^2 \frac{ K^5_5 }{ 2\pi \Big(K^t_t - K^5_5\Big)^{2}}\, ,
\end{equation}
and the radius of the extra dimensions, expressed in four-dimensional units, is given by
\begin{equation}\label{L03}
L_0 = { m_4^{-1}\  \sqrt{4 \pi ({K^t_t -K^5_5 })}} \  
\, .
\end{equation}

\section{Space variation of the four-dimensional Planck mass}

Here we discuss the variation of the four-dimensional Planck mass (and, accordingly, the four-dimensional gravitational constant $G_4 =1/{m_4}^2$) for the solution (\ref{L0n2}). In this case (f(R)=R+c, \ \n=2), the terms dropped in \eqref{TKr} have the form 
\beq
\frac{h''}{h {L_0}^{4}} \sim \frac{b''}{b {L_0}^{4}}  \sim \frac{M}{{L_0}^{4} r^3(1-2M/r)} \quad \mbox{and}  \quad \frac{{h'}^2}{h^2 {L_0}^{4}} \sim \frac{{b'}^2}{b^2 {L_0}^{4}} \sim \frac{M^2}{{L_0}^{4} r^4(1-2M/r)^2}.  
\eeq
Therefore, $\left< T^{B}_{A} \right>$ can be represented as
\bear
&& \left< T^{t}_{t} \right> = \frac{K^t_t }{{L_0}^6} +\frac{P^t_t }{{L_0}^4}\frac{M }{r^3\left( 1-2M/r \right)} +\frac{Q^t_t }{{L_0}^4}\frac{M^2}{r^4\left( 1-2M/r \right)^2} +O\left( \frac{1}{{L_0}^2 l(r)^4} \right) \, , \nn
&&  \left< T^{r}_{r} \right> = \frac{K^t_t }{{L_0}^6} +\frac{P^r_r }{{L_0}^4}\frac{M}{r^3\left( 1-2M/r \right)} +\frac{Q^r_r }{{L_0}^4}\frac{M^2 }{r^4\left( 1-2M/r \right)^2} +O\left( \frac{1}{{L_0}^2 l(r)^4} \right)\, , \nn
&& \left< T^{\theta}_{\theta} \right> = \left< T^{\varphi}_{\varphi} \right> = \frac{K^t_t }{{L_0}^6} +\frac{P^\theta_\theta }{{L_0}^4}\frac{M }{r^3\left( 1-2M/r \right)} +\frac{Q^\theta_\theta }{{L_0}^4}\frac{M^2}{r^4\left( 1-2M/r \right)^2} +O\left( \frac{1}{{L_0}^2 l(r)^4} \right)\, , \nn 
&& \left< T^{5}_{5} \right> = \left< T^{6}_{6} \right> = \frac{K^5_5 }{{L_0}^6} +\frac{P^5_5 }{{L_0}^4}\frac{M }{r^3\left( 1-2M/r \right)} +\frac{Q^5_5 }{{L_0}^4}\frac{M^2}{r^4\left( 1-2M/r \right)^2} +O\left( \frac{1}{{L_0}^2 l(r)^4} \right)\, ,
\ear
where $P^B_A$ and $Q^B_A$ are dimensionless constants. We denote
\bear
&& h(r)=1-\frac{2 M}{r} +\delta h(r), \ b(r)=\left(1-\frac{2 M}{r}\right)^{-1} +\delta b(r), \ c= \frac{2 {m_6}^2}{(K^t_t -K^5_5)^{3/2}} +\delta c, \nn
&& L(r) = \frac{(K^t_t -K^5_5)^{1/4}}{m_6} +\delta L(r), \ R(r) = \frac{2 {m_6}^2}{\sqrt{K^t_t -K^5_5}} +\delta R(r)\, .
\ear
Among the equations
\beq
\nabla_B\left< T^{B}_{A} \right> = 0
\eeq
the only nontrivial one is
\bear
\nabla_B\left< T^{B}_{r} \right> &=& \frac{1}{(K^t_t - K^5_5)} \left(2 {m_6}^3 (K^t_t - K^5_5)^{1/4}\frac{d (\delta L)}{d r}  -\frac{M (P^r_r +2 P^\theta_\theta)}{r^4 (1-2M/r)^2 } \right.
\nn && \left. - \frac{M^2 (P^t_t -P^r_r +2 Q^\theta_\theta)}{r^5 (1-2M/r)^2 } -\frac{2 M^2  Q^r_r}{r^5 (1-2M/r)^3 } -\frac{M^3 (Q^t_t - Q^r_r )}{r^6 (1-2M/r)^3 } \right)
=0 \, .
\ear
The solution of this equation is
\begin{align}
\delta L(r) {m_6}^3 &=  \frac{1}{(K^t_t -K^5_5)^{1/4}} \left[ \frac{-3P^t_t -P^r_r +4P^\theta_\theta +3Q^t_t +3Q^r_r -6Q^\theta_\theta }{32 M^2 } \ln\left( 1-\frac{2 M}{r} \right)  \right]  
\nn & 
+\frac{1}{Mr} \left[ -\left( 1 -\frac{M}{r} -\frac{2M^2}{3 r^2} \right)\frac{3(P^t_t +2 Q^\theta_\theta)}{16 (1-2M/r)}  
\right. \nn & \left.
-\left( 1 -\frac{M}{r} +\frac{2M^2}{r^2} \right) \frac{P^r_r}{16 (1-2M/r)}
+\left( 1+\frac{M}{r} \right) \frac{P^\theta_\theta}{4}
\right. \nn & \left.
+\left( 1 -\frac{3M}{r} +\frac{4M^2}{3r^2}  +\frac{2 M^3}{3r^3} \right) \frac{3 Q^t_t}{16 (1-2M/r)^2}
\right. \nn & \left.
+\left( 1 -\frac{3M}{r} +\frac{4M^2}{3r^2}  -\frac{2 M^3}{3r^3} \right) \frac{3 Q^r_r}{16 (1-2M/r)^2}
\right] \, .
\end{align}
For $r \gg 2M$
\begin{align}
\delta L(r)  & = -\frac{  (P^r_r +2P^\theta_\theta)}{6(K^t_t -K^5_5)^{1/4}}\frac{M}{{m_6}^3 r^3}\Bigg(1 +O\left( \frac{M}{r} \right)\Bigg)\, .
\end{align}
Then, the expression \eqref{m4} transforms to
\begin{align}
{m_4}^2&={m_{6}}^{4}\int d^2 y \sqrt{|g^{(2)}|} =4\pi \Big({L_0}+\delta L(r)\Big)^2 {m_{6}}^{4} 
\nn & = 4\pi  {m_{6}}^{2} \sqrt{K^t_t -K^5_5}  -\frac{4\pi}{3} \Big(P^r_r +2P^\theta_\theta \Big) \frac{M}{ r^3} \Bigg(1 +O\left( \frac{M}{r} \right)\Bigg)  .
\end{align}
One can see that the four‑dimensional Planck mass exhibits a weak dependence on the radial coordinate. Although the deviation from the conventionally constant value is expected to be small, it is worth taking this effect into account in future observations.

\section{Conclusion}

In this work, we have investigated static, spherically symmetric solutions in higher-dimensional $f(R)$ gravity with compact extra dimensions forming a sphere of constant radius.

Our main result is twofold. First, we have shown that in pure vacuum $f(R)$ gravity, the field equations admit an exact solution whose four-dimensional part is the Schwarzschild--de Sitter metric, while the extra dimensions remain stabilised. However, this solution inevitably relates the size of the internal space to the four-dimensional cosmological constant, leading to a value that is many orders of magnitude too large to match observations.

Second, we have demonstrated that the obstruction mentioned above can be overcome by including the vacuum polarisation effects of quantised matter fields nonminimally coupled to curvature. A closely analogous issue arises in the context of the cosmological constant, where the observed value can only be recovered by postulating a severe fine‑tuning of the quantum corrections. In the semiclassical approach, the quantum contribution to the stress-energy tensor can compensate the geometric part, allowing for an asymptotically flat Schwarzschild solution in four dimensions as a special case. In this regime, the size of the extra dimensions is determined by the vacuum expectation values of the quantum fields rather than by the bare parameters of the gravitational action.

We have also examined the backreaction of quantum fluctuations and found that the effective four-dimensional Planck mass acquires a weak radial dependence near the black hole, which may lead to testable deviations from general relativity in strong-field regimes.


Our results suggest that higher-dimensional $f(R)$ gravity, supplemented by quantum vacuum effects, provides a consistent framework for obtaining standard four-dimensional black hole solutions. 
Future work will be devoted to the stability analysis of these solutions, the investigation of their phenomenological signatures (in particular, the possible radial dependence of the effective Planck mass)  and their embedding into more fundamental frameworks, such as string theory.

\begin{acknowledgments}
The work of AAP was funded by the Russian Science Foundation, grant No.~25-21-00711.  The work of SGR was  funded by the Ministry of Science and Higher Education of the Russian Federation, Project ”Studying physical phenomena in the micro- and macro-world to develop future technologies”
FSWU-2026-0010
\end{acknowledgments}

\bibliography{Ru-Article_7}

\end{document}